\documentclass[aps,pre,twocolumn,showpacs]{revtex4}
\usepackage{graphicx}
\input epsf

\begin{document}

\title{The Glass Transition and Liquid-Gas Spinodal Boundaries of 
Metastable Liquids}

\author{S.S. Ashwin$^1$\footnote{E-mail: ashwinss@jncasr.ac.in\\
Present Address: Rockefeller University
1230 York Avenue, New York 10021, USA},
Gautam I. Menon$^2$\footnote{E-mail: menon@imsc.res.in} and
Srikanth Sastry$^1$\footnote {E-Mail: sastry@jncasr.ac.in}}

\affiliation{$^1$ Theoretical Sciences Unit,
Jawaharlal Nehru Centre for Advanced Scientific Research,\\ 
Jakkur PO, Bangalore 560 064,\\ India\\
$^2$The Institute of Mathematical Sciences, C.I.T Campus,
Taramani, Chennai 600 113,\\ India}

\date{\today}

\begin{abstract}

A liquid can exist under conditions of thermodynamic stability or
metastability within boundaries defined by the liquid-gas spinodal and
the glass transition line. The relationship between these boundaries
has been investigated previously using computer simulations, the
energy landscape formalism, and simplified model
calculations.  We calculate these stability boundaries
semi-analytically for a model glass forming liquid, employing accurate
liquid state theory and a first-principles approach to the glass
transition. These boundaries intersect at a finite temperature,
consistent with previous simulation-based studies.
 
\end{abstract}

\maketitle

When a substance is brought below its equilibrium freezing temperature
and yet maintained in the liquid state, it is referred to as a
supercooled liquid.  Even if crystallization is avoided a liquid
cannot be supercooled indefinitely; it transforms into an amorphous
solid {\it via} the {\em glass transition}, at a temperature determined in
experiments by the cooling rate.  The {\em ideal glass transition}
refers to the putative transformation to an amorphous solid which has
been argued to occur in the limit of infinitesimally small cooling
rates. Such a transition would define an ultimate limit beyond which a
substance cannot exist in the supercooled liquid state.

If a liquid were instead to be heated beyond the boiling point - and
the transformation into the gas phase averted - it is termed a
superheated liquid. Such a liquid cannot be superheated to arbitrarily
high temperatures since growing density fluctuations will trigger a
catastrophic transformation into the gaseous state at rates which
increase with the degree of superheating. Mean-field theory identifies
a degree of superheating for which the compressibility diverges, at
which point the liquid would spontaneously transform into a gas.  The
locus of such points defines another ultimate limit to the stability
of the liquid state, the liquid-gas spinodal \cite{debenebook}.

Clarifying the relationship between these two limits is fundamental to
a deeper understanding of the liquid state. In addition to
conventional materials in the liquid state, much interest has recently
focussed on such questions in the context of colloidal
fluids. In these systems, a unified picture of arrested states
(colloidal gels and glasses) has been the subject of considerable
investigation \cite{trappe}. The understanding of the role of the unstable region in
the phase diagram demarcated by the liquid-gas and gas-liquid
spinodals is a necessary component of a comprehensive picture of these
arrested states \cite{russel}. 

In a previous investigation \cite{sastryPRL2000} of the model liquid studied
in this letter, through computer simulations and the
energy landscape approach, it was found that the glass transition line
and the liquid-gas spinodal intersect at a finite temperature, thus
predicting a glass-gas limit of mechanical stability at
lower temperatures. More recently, similar results have been obtained
from simulations of a model of ortho-terphenyl\cite{otp} as well as
calculations based on model energy landscapes\cite{speedy,shell}.
In this letter, we describe an analysis of these limiting lines for a
realistic glass former, which uses (i) accurate liquid state theoretical
methods to calculate the equation of state \cite{ZerahHansen} of the
system, and thus the liquid-gas spinodal, coupled with (ii) a
first-principles approach to the glass transition proposed by Mezard
and Parisi \cite{parisi,mezard} to evaluate the glass transition line.
Our calculations are quantitatively in reasonable
agreement with the previous study, and also predict that the glass
transition line and the liquid-gas spinodal intersect at a finite
temperature. Our results thus provide strong support for the proposal
that such an intersection of these limiting lines is the generic
behavior for liquids.

The model system we study is the Kob-Andersen binary mixture
Lennard-Jones (KA BMLJ) liquid\cite{KobAndersenBMLJ}, consisting of an
80:20 mixture of $A$ and $B$ particles interacting via a
Lennard-Jonnes potential with parameters
$\epsilon_{AB}/\epsilon_{AA}=1.5$, $\epsilon_{BB}/\epsilon_{AA}=0.5$,
$\sigma_{AB}/\sigma_{AA}=0.8$, $\sigma_{BB}/\sigma_{AA}=0.88$. We
report energies in units of $\epsilon_{AA}$, lengths in units of
$\sigma_{AA}$, and temperatures in units of $\epsilon_{AA}/k_B$. This
system has been extensively studied as a model glass former. Our
procedure for calculating the glass transition line, which we
summarize below, is close to that employed in Ref.s
\cite{coluzziJCP2000,coluzziCondmat}.

We compute the equation of state of the system we study, KA BMLJ, from
evaluation of the $A-A$, $A-B$ and $B-B$ pair correlation functions.  Liquid
state theoretical methods provide accurate analytic tools for the
computation of such pair correlation functions and we exploit such
methods here. Our calculations use the Zerah-Hansen (ZH) closure
scheme \cite{ZerahHansen} coupled to the Ornstein-Zernike (OZ)
equation. The Zerah-Hansen closure interpolates between hyper-netted
chain (HNC) and the soft-core mean-spherical approximation (SMSA)
closures\cite{hansenMcdonald}; as described in Ref.\cite{coluzziJCP2000}, such an
interpolation reproduces thermodynamic quantities (energy and
pressure) accurately.

For the binary system, with indices $a$ and $b$ indicating 
particle types, the OZ equation is 
\begin{equation}
h_{ab}(k) = c_{ab}(k) + \sum_{i=a,b}\rho_{i}c_{ai}(k)*h_{ib}(k)
\label{eqOZ}
\end{equation}
where $\rho_{i}$ are the partial number densities, $h_{ab}$ is defined
as $g_{ab}(r)-1$ ($g_{ab}(r)$ being the $ab$ pair correlation
function) and $c_{ab}(r)$ is the direct correlation function. $h_{ab}
(k)$ {\it etc} indicate space Fourier transforms.  Defining
$\gamma_{ab}(r) = h_{ab}(r)-c_{ab}(r)$,
\begin{equation}
\gamma_{ab}(k) = \sum_{i=a,b}\rho_{i}c_{ai}(k)(c_{ib}(k)+\gamma_{ib}(k)).
\end{equation}

The ZH closure is of the form
\begin{eqnarray}
c^{a}c^{b}\rho^{2}g_{ab}(r) & = & \exp(-\beta v_{R}^{ab}(r)) \\ \nonumber 
& \times & \left(1 +\frac{exp[f^{ab}(r)[\gamma_{ab}(r)-\beta v_{A}^{ab}(r)]-1}{f^{ab}(r)}\right)
\label{eqZH}
\end{eqnarray}
where $\beta \equiv 1/k_B T$, $c^{a}$ and $c^{b}$ are the concentrations of  $A$  type and $B$ type particles, respectively, $v_{R}$ and $v_{A}$ are attractive and
repulsive contributions of the pairwise potential $v$, and are given by $v_{R} =
v(r) - v_{min} ~~ (r \le r_{min})$ and $ v_{R} = 0 ~~ (r \geq
r_{min})$; $v_{A} = v_{min} ~~ (r \le r_{min})$ and $ v_{R} = v(r) ~~
(r \geq r_{min})$. The function $f$ provides an interpolation between
the SMSA and the HNC: $f^{ab}= 1 -exp\left[
-\frac{r}{\sigma_{ab}\alpha}\right]$, with $\alpha$ governing the
switch between HNC and SMSA.  The parameter $\alpha$ is typically
chosen by demanding thermodynamic consistency between equations of
states obtained by the virial and compressibility routes. Other
choices, however, are possible \cite{coluzziJCP2000}, and here we choose the parameter
$\alpha$ by comparison with the simulation data for the studied system
in a range of temperature and density values.  We find, as expected,
that the value of $\alpha$ that yields the best match of energy and
pressure values with those from simulations varies both with
temperature $T$ and density $\rho$. Based on the estimation of such
$\alpha$ values for a small set of temperature and density values, we
choose a linear functional form for the $\beta$ and $\rho$ dependence
of $\alpha$, $\alpha(\rho, T) = 11.17 -11.7\rho + 6.65\beta$, which 
allows us to obtain quantitatively good results for thermodynamic quantities. 

We solve the OZ equation, along with the ZH closure, for $\gamma_{ab}$
using the procedure outlined in Ref. \cite{mohamad}.  Pair correlation
functions are obtained first for low density and high temperatures.
The pair correlation functions so obtained are used as initial guesses
for the iterative solution of the OZ and ZH equations at lower
temperatures and higher densities.  We require iterative convergence
at the level of $10^{-9}$ in $\sum_{i}(g_{n+1}(r_i)-g_{n}(r_i))^2$,
with $n$ being the iteration number and $i$ the real space mesh index.

In Fig. 1, we compare isotherms obtained using this scheme for
temperatures $T=0.4,0.5,0.6$ with the isotherms derived, in
\cite{sastryPRL2000}, from an empirical equation of state based on
simulation data. Data shown in Fig. 1 demonstrate that the equation of
state from the present calculations are in good quantitative agreement
with the results in \cite{sastryPRL2000}. Figure 2 shows the potential
energies ($e_{pot}$) obtained from the present calculations as a
function of $T$ and $\rho$, along with available values from computer
simulations \cite{sastryPRL2000}, demonstrating excellent agreement
with the simulation results.  The comparisons in Fig.s 1 and 2 
indicate that the pressure and energy values obtained from the 
present calculations are quantitatively quite accurate. 

In Fig. 3, we show the isotherms obtained for temperatures in the
range $T=0.4-0.9$. We obtain the location of the spinodal at each
temperature from the condition $\frac{\partial P}{\partial\rho}=0$.

\begin{figure}[h] 
\includegraphics[width=80mm]{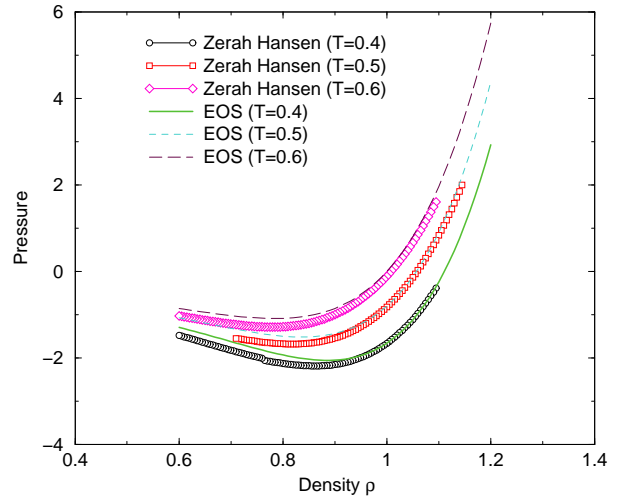}
\caption{Pressure {\it vs} density isotherms obtained
using the Zerah-Hansen scheme (symbols), compared with
the equation of state obtained from fits to the
simulation data (lines) for temperatures $T=0.4, 0.5$
and $0.6$.}
\label{comparison} 
\end{figure}

\begin{figure}[h]
\includegraphics[width=80mm]{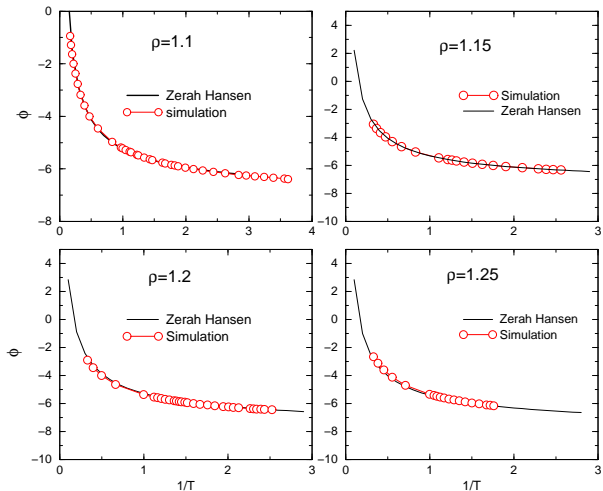}
\caption{Comparison of potential energy obtained using the
Zerah-Hansen scheme (lines) with data from molecular dynamics
simulation, shown for densities $\rho=1.1,1.15,1.2,1.25$ as a function
of the inverse temperature $\beta$.  }
\label{sliq}
\end{figure}

\begin{figure}[h]
\includegraphics[width=80mm]{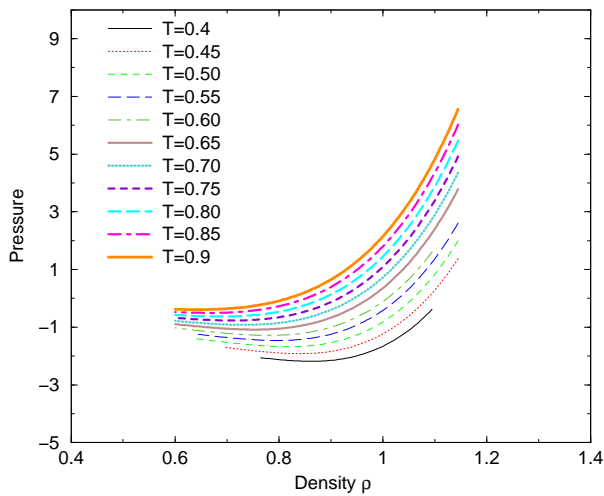}
\caption{Pressure {\it vs.} density isotherms obtained using the
Zerah-Hansen scheme for temperatures varying from $T=0.4$ to
$T=0.9$. The spinodal lines are obtained from the location of the
minima along these isotherms.  }
\label{sliq} 
\end{figure}

The glass transition temperatures at each density are calculated using
approach proposed by Mezard and Parisi \cite{parisi,mezard,monPRL,sfranz}, which has
been employed to calculate the glass transition line for the KA BMLJ
in \cite{coluzziJCP2000}. In this calculation, assuming the existence
of an exponentially large number of free energy minima whose
degeneracy determines the configurational entropy or complexity, the
thermodynamics of a system of $m$ copies of the system which are
restricted to be in the same free energy minimum is considered.  With
the replicas remaining confined close to each other, the free energy
can be written in terms of their center of mass coordinates and the
excursions of individual replicas from the center of mass. Evaluation
of the contribution from the latter is done within approximation
schemes pertaining to the Hessian or potential energy second
derivative matrix. The configurational entropy is obtained by
differentiating the resulting free energy per replica with respect to
their number, in the limit of $m \rightarrow 1$. The resulting
expression for the configurational entropy is

\begin{equation} 
S_{conf} = S_{liq}-S_{glass}
\end{equation} 

where $S_{liq}$ is the entropy of the liquid, and 
\begin{equation}
S_{glass}(\beta) = \frac{3N}{2}(1+log(2\pi)) - \frac{1}{2}<Tr(log(\beta M))>,
\end{equation}
where $M$ is the Hessian matrix of the form $M_{(i\mu),(j\nu)}
=\delta_{ij}\sum_{k}v_{\mu\nu}(r_i-r_k)-v_{\mu\nu}(r_i -r_j)$ with
$v_{\mu\nu}=\frac{\partial^2 v}{\partial r_{\mu}\partial
r_{\nu}}$. The indices $\mu$ and $\nu$ run over the coordinates and
$i,j$ and $k$ run over particle indices. The trace is calculated using
the harmonic resummation scheme, details of which (and of the discussion
above) can be found in \cite{parisi,mezard,coluzziJCP2000}. As in
\cite{sastryPRL2000,SciortinoPRL1999}, the ideal glass transition
$T_K$ is located with the condition $S_{conf} (T_K) = 0$.

\begin{figure}[h] \centering
\includegraphics[width=80mm]{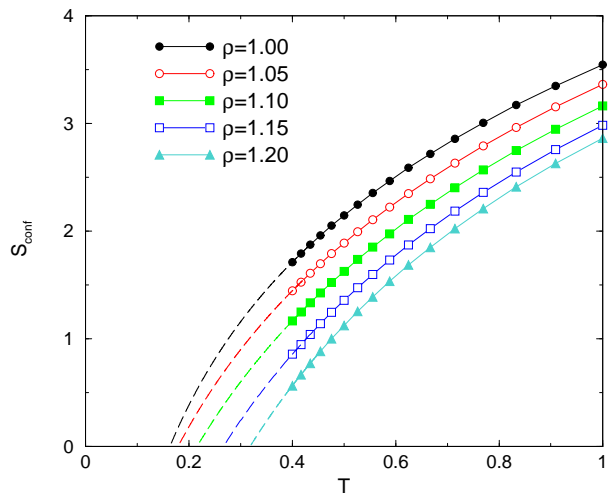}
\caption{Configurational entropy $S_{conf}$ for densities
$\rho=1.0,1.05,1.10,1.15,1.20$ are shown.  For low temperatures the
$S_{conf}$ is obtained via an extrapolation shown by the dashed
lines. The Kauzmann temperature is identified with the temperature at
which the configurational entropy vanishes.} \label{Complexity for
densities}
\end{figure} 

The entropy of the liquid is obtained from the internal energy
($e(\beta) = \frac{3k_BT}{2} + e_{pot}$) {\it via}
\begin{equation}
S_{liq}(\beta) = S^{o}_{liq} + \beta e_{liq} -\int_{0}^{\beta}d\beta^{'}e(\beta^{'}).
\end{equation}
Here $S^{o}_{liq}$ is the entropy of the perfect gas
in the infinite temperature limit and is given by
$S^{o}_{liq} = 1 -log(\rho) -c^{a}log(c^{a}) - c^{b}log(c^{b})$
where $c^{a}$ is the concentration of $A$ particles and
$c^{b}$ is the concentration of $B$ type particles.

Both $S_{liq}$ and $S_{glass}$ require knowledge of the liquid state
pair correlation functions. However, at low temperatures the liquid
integral equations do not converge, and hence extrapolations (of the
form $aT^{-2/5}+b$ for $S_{liq}$ and $a^{'} + b^{'} log(T)$ for
$S_{glass}$, based respectively on theoretical predictions for dense
liquids \cite{tarazona} and the harmonic approximation to the glass
free energy) are used to obtain $S_{conf}$ at low temperatures.
Figure 4 shows the configurational entropy {\it vs.} $T$ for
$\rho=1.0$,$1.05$,$1.1$,$1.15$,$1.20$.

\begin{figure}[h]
\centering
\includegraphics[width=80mm]{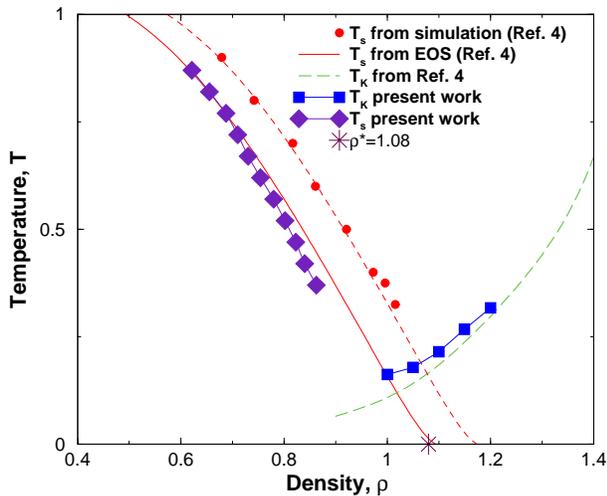}
\caption{Limits of stability of the liquid phase from present work --
spinodal labeled $T_s$ and glass transition labeled $T_K$ -- compared
with previous work \cite{sastryPRL2000}: (circles) spinodal estimated
from inverse compressibility (dashed line is a guide to the eye),
(solid line) from the empirical equation of state, and (long dashed
line) glass transition line from energy landscape approach. The
present calculation of the spinodal agrees quite well with that from
the empirical equation of state, while the glass transition
temperatures are seen to be slightly overestimated. The two limiting
lines intersect at $\rho \sim 0.95$, as compared to $\rho \sim 1.08$
from earlier work. (The intersection density in \cite{sastryPRL2000} is
$1.02$ if the spinodal estimate from the empirical equation of state
is used instead of the simulation estimates, a number closer to the one obtained from our present calculations.)}
\label{phasediag} 
\end{figure}

The liquid-gas spinodal and the glass transition lines obtained are
shown in Fig. 5, along with results from earlier work
\cite{sastryPRL2000}. Our results reproduce the features observed in
earlier work both qualitatively and, to a good extent, quantitatively,
thus providing support for the general applicability of this
approach. The calculated spinodal temperatures are in reasonable
agreement with the ones estimated earlier from the empirical free
energy. The glass transition temperatures obtained here are slightly
higher than those estimated in previous work; this may be a
consequence of the approximations involved in evaluating the glass
entropies in the present case.

Together with previous work
\cite{sastryPRL2000,otp,speedy,shell}, these results lend strong
support to the scenario that the liquid-gas spinodal and the glass
transition line should typically intersect at a finite low
temperature. Unlike previous work, where the results were based either
on simulations, or on assumed model energy landscapes, our
calculations describe a realistic model glass former, whose properties
are evaluated within a self-contained and accurate framework. We believe 
that the approach used here and its 
extensions may be useful in understanding arrested states in 
contexts similar to the one we study. 

The authors acknowledge support from the DST (India).    


\begin{references}


\bibitem{debenebook} P. G. Debenedetti, {\it Metastable Liquids:
Concepts and Principles} (Princeton University Press, Princeton,
N. J. 1996).

\bibitem{trappe} V. Trappe and P. Sandk\"uler, {\it Current Opinion in Colloid and Interface Science} {\bf 8} 494 (2004). 

\bibitem{russel} M. C. Grant and W. B. Russel, {\it Phys. Rev. E} {\bf 47} 2606 (1993). 


\bibitem{sastryPRL2000} S.Sastry, {\it Phy. Rev. Lett.}, {\bf 85}, 590, (2000).

\bibitem{otp} E. La Nave, S. Mossa, F. Sciortino, and P. Tartaglia,
{\it J. Chem. Phys.}, {\bf 120}, 6128 (2004). 

\bibitem{speedy} R.J. Speedy, {\it J. Phys. Cond. Matt}, {\bf 15}, S1243 (2003).

\bibitem{shell} M.S. Shell and P.G. Debenedetti, {\it Phys. Rev. E},
{\bf 69}, 051102 (2004).

\bibitem{KobAndersenBMLJ} W. Kob and H.C. Andersen, {\it Phy. Rev. E.}, {\bf 51}, 4626, (1995). 

\bibitem{ZerahHansen}G. Zerah and J.P. Hansen, {\it J. Chem. Phys.}, {\bf 84}, 2336, (1986).

\bibitem{hansenMcdonald}J.P. Hansen and I. R. McDonald, {\it Theory of Simple Liquids}, Academic Press, New York (1986).

\bibitem{parisi}M. Mezard and G. Parisi, {\it Phy. Rev. Lett.}, {\bf 82}, 747, (1999).

\bibitem{mezard} M. Mezard and G. Parisi, {\it J. Chem. Phys.} {\bf 111} 1076 (1999). 

\bibitem{coluzziJCP2000} B. Coluzzi, G. Parisi, P. Verrochio, {\it J. Chem. Phys.}, {\bf 112}, 2933, (2000).

\bibitem{coluzziCondmat} B. Coluzzi, G. Parisi, P. Verrochio, {\it cond-mat/0007144}.

\bibitem{mohamad} M.D. Lotfollahi, H. Modarress,{\it J. Chem. Phys.}, {\bf 116}, 2487, (2002).

\bibitem{monPRL} R. Monasson, {\it Phys. Rev. Lett.} {\bf 75},
2847 (1995).

\bibitem{sfranz} S. Franz and G. Parisi, {\it J. Phys. I (France)
} {\bf 5}, 1401 (1995).

\bibitem{SciortinoPRL1999} F. Sciortino, W. Kob and P. Tartaglia, 
{\it Phy. Rev. Lett.}, {\bf 83}, 3214, (1999).

\bibitem{tarazona} Y. Rosenfeld and P. Tarazona, {\it Mol. Phys.} {bf 84} 141 (1998).


\end{references}
\end{document}